\documentclass[aps,prb,twocolumn,superscriptaddress]{revtex4-2}

\usepackage{graphicx}
\usepackage{amsmath}
\usepackage{textcomp}
\usepackage{array}[=2016-10-06]
\usepackage{booktabs}
\usepackage{multirow}
\usepackage{color}
\usepackage[normalem]{ulem}

\usepackage{hyperref}

\newcommand{\matonly}[1]{
}

\begin{document}

\title{
Modified anisotropy and distinct phase diagrams in the nearly isotropic $S=1$ triangular-lattice antiferromagnet $\mathrm{Rb_2Ni(SeO_3)_2}$
}

\author{Xuejuan Gui}
\affiliation{Laboratory for Neutron Scattering, School of Physics, Renmin University of China, Beijing 100872, China}
\affiliation{Key Laboratory of Quantum State Construction and Manipulation (Ministry of Education), Renmin University of China, Beijing, 100872, China}

\author{Xu-Guang Zhou}
\affiliation{High Magnetic Field Laboratory, Hefei Institutes of Physical Science, Chinese Academy of Sciences, Hefei 230031, Anhui, China}
\affiliation{Institute for Solid State Physics, University of Tokyo, Kashiwa, Chiba 277-8581, Japan}

\author{Chen Zhang}
\affiliation{Department of Physics, Southern University of Science and Technology, Shenzhen 518055, China}
\affiliation{School of Physical Sciences, Great Bay University and Great Bay Institute for Advanced Study, Dongguan 523000, China}

\author{Zhongcen Sun}
\author{Yuan Sun}
\author{Chenglin Shang}
\affiliation{Laboratory for Neutron Scattering, School of Physics, Renmin University of China, Beijing 100872, China}
\affiliation{Key Laboratory of Quantum State Construction and Manipulation (Ministry of Education), Renmin University of China, Beijing, 100872, China}

\author{Akira Matsuo}
\author{Yasuhiro H. Matsuda}
\author{Koichi Kindo}
\affiliation{Institute for Solid State Physics, University of Tokyo, Kashiwa, Chiba 277-8581, Japan}

\author{Henrik M. Rønnow}
\affiliation{Laboratory for Quantum Magnetism, Institute of Physics, École Polytechnique Fédérale de Lausanne (EPFL), CH-1015 Lausanne, Switzerland}

\author{Weiqiang Yu}
\author{Rong Yu}
\affiliation{Key Laboratory of Quantum State Construction and Manipulation (Ministry of Education), Renmin University of China, Beijing, 100872, China}

\author{Peng Cheng}
\affiliation{Laboratory for Neutron Scattering, School of Physics, Renmin University of China, Beijing 100872, China}
\affiliation{Key Laboratory of Quantum State Construction and Manipulation (Ministry of Education), Renmin University of China, Beijing, 100872, China}

\author{Juanjuan Liu}
\email{juanjuanliu@ruc.edu.cn}
\affiliation{Laboratory for Neutron Scattering, School of Physics, Renmin University of China, Beijing 100872, China}
\affiliation{Key Laboratory of Quantum State Construction and Manipulation (Ministry of Education), Renmin University of China, Beijing, 100872, China}

\author{Jieming Sheng}
\email{shengjm@gbu.edu.cn}
\affiliation{School of Physical Sciences, Great Bay University and Great Bay Institute for Advanced Study, Dongguan 523000, China}

\author{Jinchen Wang}
\email{jcwang\_phys@ruc.edu.cn}
\affiliation{Laboratory for Neutron Scattering, School of Physics, Renmin University of China, Beijing 100872, China}
\affiliation{Key Laboratory of Quantum State Construction and Manipulation (Ministry of Education), Renmin University of China, Beijing, 100872, China}
\affiliation{Laboratory for Quantum Magnetism, Institute of Physics, École Polytechnique Fédérale de Lausanne (EPFL), CH-1015 Lausanne, Switzerland}
\affiliation{PSI Center for Neutron and Muon Sciences, 5232 Villigen PSI, Switzerland}

\begin{abstract}
The triangular-lattice antiferromagnet serves as a benchmark platform for realizing exotic quantum phases, in which magnetic anisotropy plays an essential role. 
In this work, we report modified magnetic anisotropy and phase diagrams in the nearly isotropic $S=1$ triangular-lattice antiferromagnet $\mathrm{Rb_2Ni(SeO_3)_2}$, compared with those of its isostructural analog $\mathrm{K_2Ni(SeO_3)_2}$. 
The magnetic susceptibility of $\mathrm{Rb_2Ni(SeO_3)_2}$ exhibits an easy-axis anisotropy with $g_c > g_{ab}$, opposite to the easy-plane anisotropy observed in $\mathrm{K_2Ni(SeO_3)_2}$. 
The single N\'eel transition observed at zero field splits into two transitions at around 10~T, and a magnetization plateau is observed between 21~T and 37~T for both out-of-plane ($H\parallel c$) and in-plane ($H\parallel ab$) field directions in $\mathrm{Rb_2Ni(SeO_3)_2}$. 
We construct magnetic phase diagrams that are distinct between $\mathrm{Rb_2Ni(SeO_3)_2}$ and $\mathrm{K_2Ni(SeO_3)_2}$. These findings highlight the sensitivity of field-induced magnetic phase behavior to anisotropy in nearly isotropic triangular-lattice antiferromagnets.
\end{abstract}

\pacs{ }

\maketitle

\section{Introduction}
Exploring novel quantum phases is a central pursuit in condensed matter physics.
The triangular-lattice antiferromagnet (TLAF) provides a paradigmatic frustrated system, in which the local preference for antiparallel alignment of neighboring spins cannot be simultaneously satisfied~\cite{Collins1997,Starykh2015}.
This geometric frustration produces a manifold of nearly degenerate configurations and strong quantum fluctuations, providing fertile ground for exotic states.
A variety of intriguing phases have been proposed in TLAFs as a result of the interplay among interaction parameters. 
One prominent example is the emergence of a quantum spin-liquid phase~\cite{Balents2010} driven by the competition between next-nearest-neighbor and nearest-neighbor interactions~\cite{Hu2015,Zhu2015,Iqbal2016}.
In addition, tuning the anisotropy between the transverse and longitudinal exchange couplings in the XXZ model yields a rich sequence of field-induced states, including the Y, up-up-down (UUD), V, and Umbrella phases~\cite{Yamamoto2014,Sellmann2015,Marmorini2016}.

Despite these rich theoretical proposals, their experimental realization requires stringent control of both lattice geometry and the form of the interactions. 
Recently, the XXZ model has been realized in the $S=1/2$ TLAF $\mathrm{A_2Co(SeO_3)_2}$ (A = K, Rb)~\cite{Zhong2020} and $\mathrm{Na_2BaCo(PO_4)_2}$~\cite{Zhong2019}. 
In these materials, the exchange interaction takes the easy-axis form. Spin-supersolid phases~\cite{Sheng2022,Gao2022,Xiang2024,Zhu2024,Chen2026,Cui2026}, giant magnetocaloric effects~\cite{Xiang2024} and continuous spin excitations~\cite{Sheng2025,Zhu2024,Zhu2025,Chen2026} have been observed. 
In the easy-plane regime, $\mathrm{Ba_3CoSb_2O_9}$ is the most established $S=1/2$ realization and has attracted sustained attention~\cite{Shirata2012,Zhou2012,Susuki2013,Quirion2015,Koutroulakis2015,Ma2016,Ito2017,Kamiya2018,Macdougal2020}. 
More recently, $\mathrm{K_2Ni(SeO_3)_2}$ has also been proposed as an easy-plane realization of a $S=1$ TLAF~\cite{Li2023,Kim2026}.

In systems discussed above, the anisotropy is from exchange interactions, whereas the single-ion anisotropy is considered negligible. 
However, single-ion anisotropy can also play a significant role in stabilizing quantum phases. 
This was recently demonstrated in $S=1$ TLAF $\mathrm{Na_2BaNi(PO_4)_2}$, where the two-magnon bound state and spin-nematic phases~\cite{sheng2025bose,ShengPRL2025,Huang2025} are stabilized by an easy-axis single-ion anisotropy $D$ that is four times the exchange interaction $J$. 
This raises the question whether novel phases can be realized by combining different sources of anisotropy, which may act cooperatively or compete with each other.

In this paper, we report on the $S=1$ TLAF $\mathrm{Rb_2Ni(SeO_3)_2}$. 
By combining magnetic susceptibility, specific heat, and pulsed high-field magnetization measurements, we construct magnetic phase diagrams that are distinct from the sibling compound $\mathrm{K_2Ni(SeO_3)_2}$. 
In particular, we identify a split of $T_\mathrm{N}$ around 10~T and magnetization plateaus between 21~T and 37~T in both out-of-plane and in-plane field orientations.
In addition, a modification of the magnetic anisotropy from the easy-plane type ($g_{ab}>g_c$) in $\mathrm{K_2Ni(SeO_3)_2}$ to the easy-axis type ($g_{c} > g_{ab}$) in $\mathrm{Rb_2Ni(SeO_3)_2}$ is observed.
This contrasts with the expectation that nonmagnetic interlayer spacer ions have little influence on magnetic properties.
We suggest that the interplay between single-ion and exchange anisotropies warrants further investigation. Tuning the competition between these anisotropic terms could provide a promising route toward novel magnetic phases.

\section{Methods}
$\mathrm{Rb_2Ni(SeO_3)_2}$ and $\mathrm{K_2Ni(SeO_3)_2}$ single crystals used in this study were grown using solid-state reaction methods reported in Refs.~\cite{Zhong2020,Li2023}.
For $\mathrm{Rb_2Ni(SeO_3)_2}$, high-purity starting materials, Rb$_2$CO$_3$ (99.99\%), NiO (99.99\%), and SeO$_2$ (99.99\%), were weighed in a molar ratio of 1.2:1:2.2, with a total mass of 1.5~g.
The mixture was ground to ensure uniformity and then transferred into a quartz tube, which was sealed under a vacuum of $10^{-3}$~Pa.
The sealed tube was heated to 750\textdegree C and held at this temperature for 1600~min.
It was then cooled to 200\textdegree C at a rate of 5\textdegree C/h, followed by furnace cooling to room temperature.
After washing with deionized water, yellow, transparent single crystals were obtained, as shown in Fig.~\ref{fig1}(a).

The crystal structure of Rb$_2$Ni(SeO$_3$)$_2$ was determined by single-crystal (four-circle) X-ray diffraction at 300~K.
Data were collected using a Bruker D8 VENTURE diffractometer equipped with a PHOTON III C14 detector and Mo K$\alpha$ radiation ($\lambda = 0.71073$~\AA) with a graphite monochromator.
Unit-cell refinement, data integration, and absorption correction were performed using the Bruker APEX3 software package.
The structure was solved by intrinsic phasing using SHELXT and refined by full-matrix least squares on $F^2$ using SHELXL as implemented in Olex2~\cite{Sheldrick2015,Dolomanov2009}.
Analysis with PLATON revealed no evidence of higher symmetry~\cite{Spek2009}.
X-ray diffraction from the $(00l)$ lattice planes was measured using a Bruker D8 ADVANCE powder diffractometer with Cu K$\alpha$ radiation.

Magnetic susceptibility and specific-heat measurements were carried out using a Quantum Design Magnetic Property Measurement System (MPMS) and a Physical Property Measurement System (PPMS), respectively.
In addition, pulsed high-field magnetization measurements with peak fields of 26, 31, 41, and 60~T were performed at the International MegaGauss Science Laboratory, University of Tokyo, using a nondestructive short-pulse magnet system.

\section{Results}
\begin{figure}[b]
\includegraphics [width=\columnwidth]{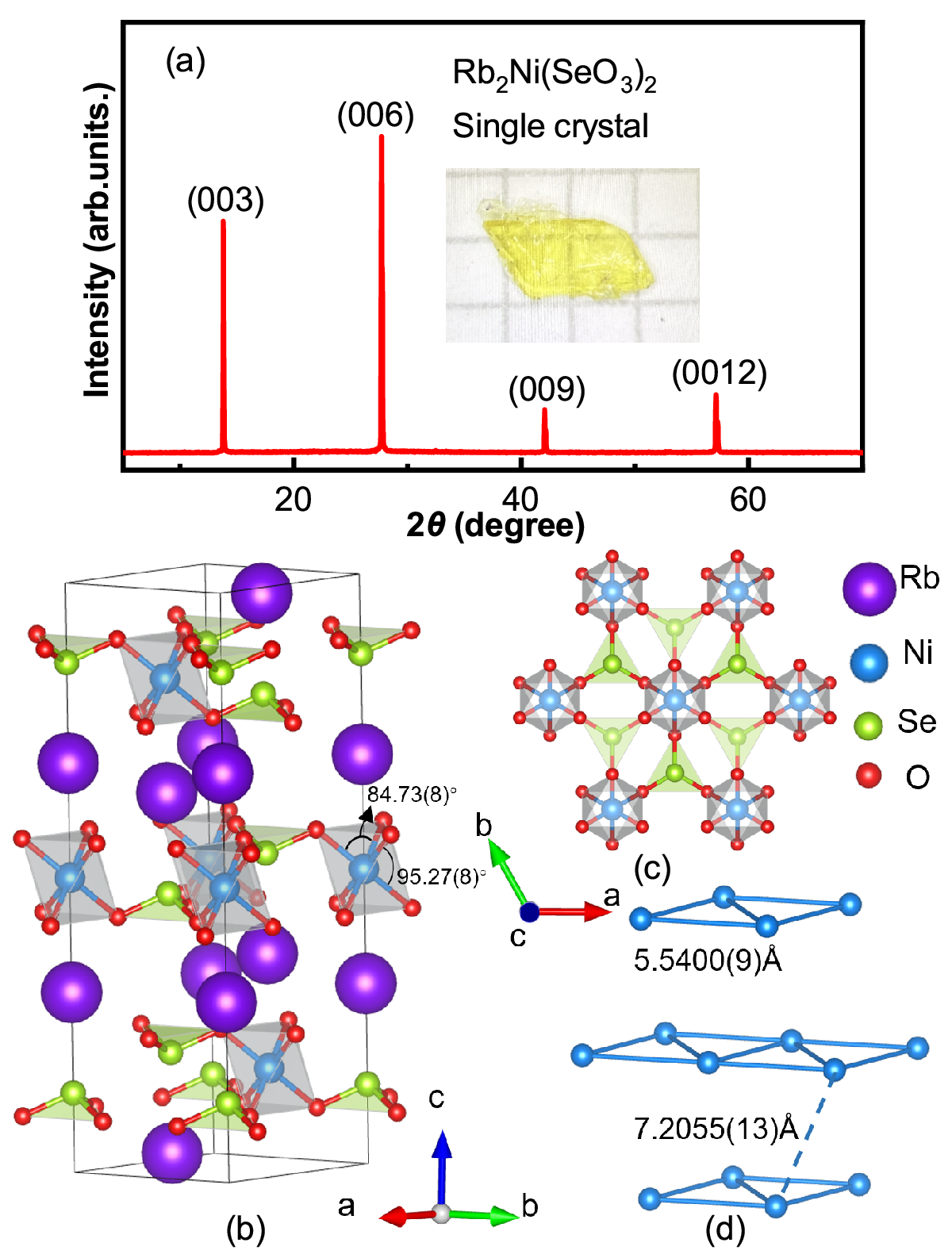}
\caption{
(a) X-ray diffraction of the $(00l)$ lattice planes. The inset shows a single crystal on millimeter-grid paper.
(b--d) Crystal structure of $\mathrm{Rb_2Ni(SeO_3)_2}$ and its triangular network. The ions, bond angles of the NiO$_6$ octahedra, and intra- and interlayer Ni--Ni bond distances are indicated.
}
\label{fig1}
\end{figure}

$\mathrm{Rb_2Ni(SeO_3)_2}$ crystallizes in the trigonal space group $R\bar{3}m$ (No.~166), with lattice parameters $a=b=5.5400(8)$~\AA\ and $c=19.370(4)$~\AA, and is isostructural with the related compound $\mathrm{K_2Ni(SeO_3)_2}$~\cite{Li2023}.
The platy crystals have the flat surface that is perpendicular to the $c$ axis  and produces the $(00l)$ reflections shown in Fig.~\ref{fig1}(a).
The structural parameters refined from single-crystal X-ray diffraction are listed in Table~\ref{table:crystal_structure}.

\begin{table}[t]
\centering
\caption{
Structure details for $\mathrm{Rb_2Ni(SeO_3)_2}$ 
determined by single-crystal X-ray diffraction at 300~K. 
The compound crystallizes in the trigonal space group $R\bar{3}m$ (No.~166)
with refined lattice parameters 
$a = 5.5400(8)$~\AA, 
$b = 5.5400(8)$~\AA, 
$c = 19.370(4)$~\AA,
$\alpha = 90^\circ$, 
$\beta = 90^\circ$, 
$\gamma = 120^\circ$.
A total of 3612 reflections and 187 unique reflections were collected.
The refinement converged to
$R_1 = 0.0167$ and 
$wR_2 = 0.0397$ for all data.
All atomic sites are fully occupied.
}
\label{table:crystal_structure}
\begin{ruledtabular}
\small
\begin{tabular*}{\columnwidth}{@{\extracolsep{\fill}}cccccc}
Atom & Wyck. & $x$ & $y$ & $z$ & $U_{\mathrm{eq}}$ \\
\hline
Rb1 & 6c  & 0.33333 & 0.66667 & 0.36122(2) & 0.0148(2) \\
Ni1 & 3b  & 0.33333 & 0.66667 & 0.16667 & 0.0075(2) \\
Se1 & 6c  & 0.66667 & 0.33333 & 0.19864(2) & 0.0073(2) \\
O1  & 18h & 0.50596(16) & 0.49404(16) & 0.23571(11) & 0.0121(4) \\
\end{tabular*}
\end{ruledtabular}
\end{table}

As illustrated in Fig.~\ref{fig1}(b--d), the magnetic Ni$^{2+}$ ion is coordinated by six oxygen atoms, forming a local $\mathrm{NiO}_6$ octahedron. The octahedral crystal field acting on the 3d$^8$ electrons of Ni$^{2+}$ stabilizes the $S=1$ spin state~\cite{Abragam1970}. Magnetic anisotropy can arise from octahedral distortions and spin--orbit coupling (SOC). These octahedra, interconnected by $\mathrm{SeO}_3$ pyramids, form an equilateral triangular lattice in the $ab$ plane. Along the $c$ axis, the triangular layers follow an ABC stacking sequence and are separated by nonmagnetic Rb$^+$ ions. 
Together with the relatively large interlayer spacing, this gives the system a quasi-two-dimensional structure, with interlayer magnetic interactions expected to be much weaker than intralayer interactions.

\begin{figure}[tb]
\includegraphics [width=\columnwidth]{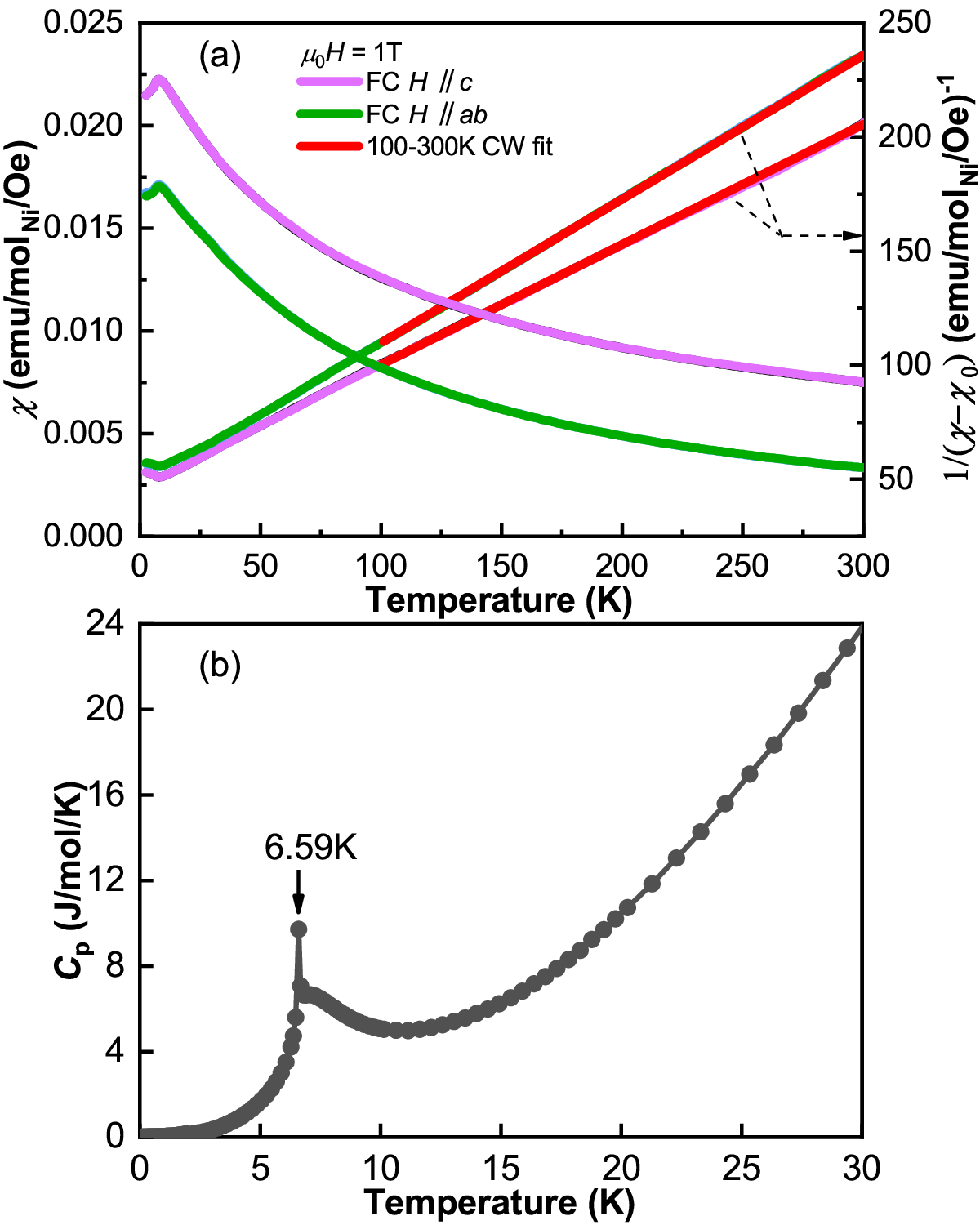}
\caption{(a) Temperature dependence of the magnetic susceptibility $\chi$ (left axis) and the inverse susceptibility $1/(\chi-\chi_0)$ (right axis) for magnetic fields applied parallel to the $ab$ plane and the $c$ axis. Here, $\chi_0$ denotes the temperature-independent contribution. The red solid lines represent Curie--Weiss fits performed over the temperature range 100--300~K. Zero-field-cooled (ZFC) and field-cooled (FC) measurements yield identical results; therefore, only the FC data are shown.
(b) Zero-field specific heat of $\mathrm{Rb_2Ni(SeO_3)_2}$. Arrows mark the transition temperature.
}
\label{fig2}
\end{figure}

Figure~\ref{fig2}(a) shows the temperature dependence of the magnetic susceptibility, $\chi(T)$, measured under magnetic fields applied parallel to the $ab$ plane and the $c$ axis. 
For both field directions, the susceptibility exhibits Curie--Weiss behavior at high temperatures and an antiferromagnetic (AFM) transition at the N\'eel temperature $T_\mathrm{N} = 6.6$~K. 
Notably, the susceptibility is larger for $H\parallel c$ in $\mathrm{Rb_2Ni(SeO_3)_2}$. 
This anisotropy is opposite to that observed in $\mathrm{K_2Ni(SeO_3)_2}$, where the susceptibility for $H\parallel ab$ is larger~\cite{Li2023}. 
The AFM transition is also confirmed by the specific heat shown in Fig.~\ref{fig2}(b). 
In addition, a broad hump in specific heat appears slightly above $T_\mathrm{N}$, suggesting the presence of short-range magnetic correlations, possibly originating from frustration or the quasi-2D nature of magnetism.

Curie--Weiss fits, performed after subtracting the temperature-independent contribution, yield a Curie--Weiss temperature $\theta^{ab}_{\mathrm{CW}} = -74.1$~K and an effective magnetic moment $\mu^{ab}_{\mathrm{eff}} = 3.56~\mu_{\mathrm{B}}/\mathrm{Ni}$ for $H\parallel ab$, and $\theta^{c}_{\mathrm{CW}} = -90.2$~K and $\mu^{c}_{\mathrm{eff}} = 3.89~\mu_{\mathrm{B}}/\mathrm{Ni}$ for $H\parallel c$. 
Assuming $S=1$ for Ni$^{2+}$, these values correspond to $g_{ab} = 2.52$ and $g_c = 2.75$. 
The negative Curie--Weiss temperatures indicate dominant antiferromagnetic interactions. 
Using $\theta_{\mathrm{avg}}=-zJS(S+1)/3k_{\mathrm{B}}$, where $\theta_{\mathrm{avg}} = (2\theta^{ab}_{\mathrm{CW}}+\theta^{c}_{\mathrm{CW}})/3$ and $z=6$ is the triangular-lattice coordination number, we estimate the AFM exchange interaction of $J = 1.7$~meV. 
The ratio between the Curie--Weiss temperature and $T_\mathrm{N}$ yields the frustration parameter $f = |\theta_{\mathrm{CW}}|/T_\mathrm{N}$ of approximately 11 for $H\parallel ab$ and 14 for $H\parallel c$, indicating strong frustration.

\begin{figure}[tb]
\includegraphics[width=\columnwidth]{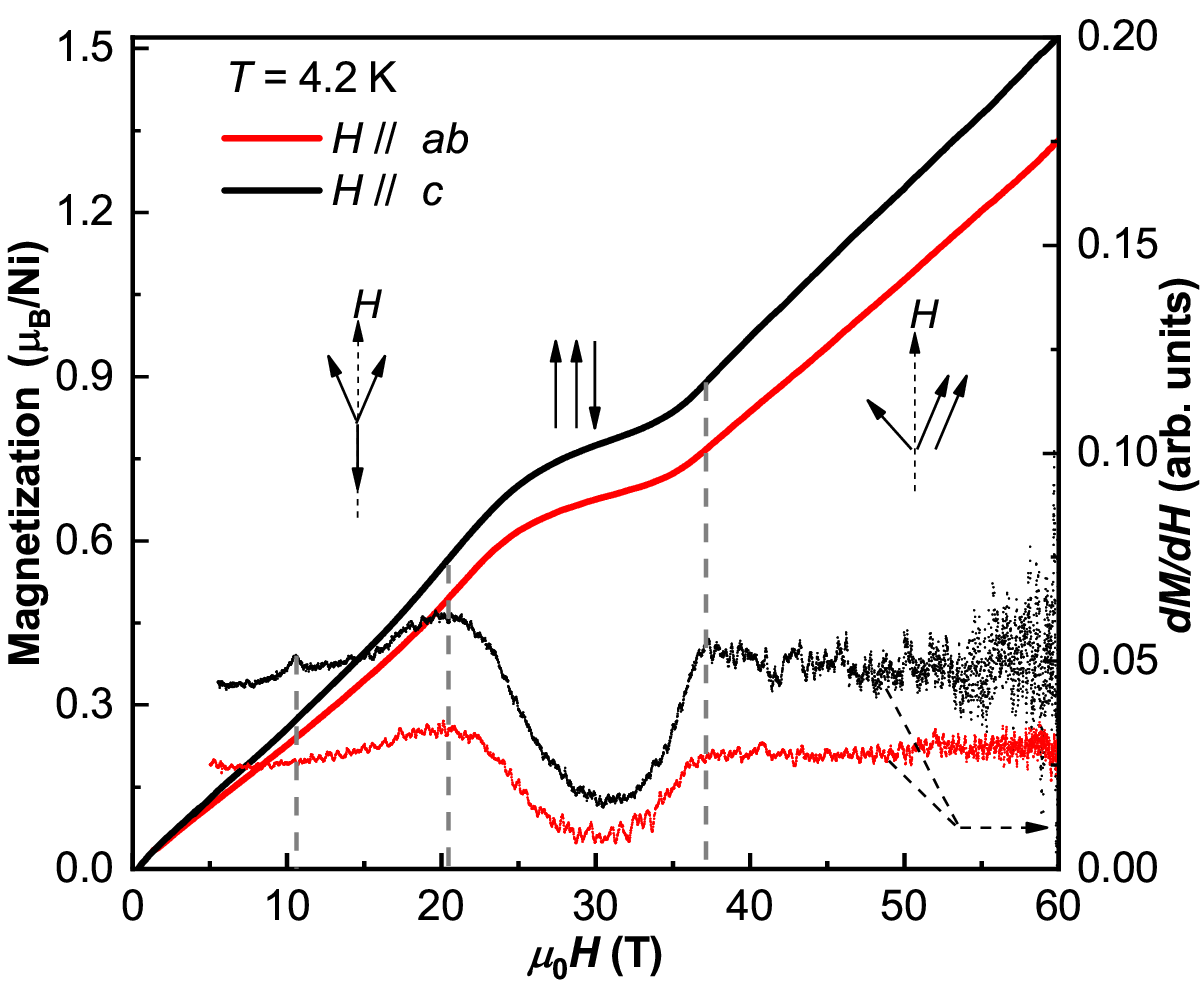}
\caption{ Pulsed high magnetic field magnetization (left axis) and $dM/dH$ (right axis) measurement at 4.2~K for field applied in the $ab$ (red) and $c$ (black) directions. Measurements on the up- and down-sweeps agree thus only the up-sweep curves are shown.
The absolute magnetization values were calibrated using 0--14~T PPMS results. Low-field $dM/dH$ data are omitted because of substantial measurement noise. Dashed lines mark the critical fields. Schematic three-sublattice spin configurations corresponding to Y, UUD and V phases are shown.
}
\label{fig3}
\end{figure}

We measured magnetization at $T = 4.2~\mathrm{K}$ using pulsed magnetic fields with various pulsed up to 60~T, applied within the $ab$ plane and along the $c$ axis, as shown in Fig.~\ref{fig3}. 
Very similar magnetic responses, including clear magnetization plateaus, are observed for the two field directions.
The lower and upper plateau boundaries, extracted from the field-derivative curves $dM/dH$, are approximately $H_{c2}\sim21$~T and $H_{c3}\sim37$~T, respectively.
These critical fields are close for both field orientations.
The saturation fields lie beyond the measurement range and are extrapolated to be approximately 86~T.
Interestingly, a weak but discernible anomaly is observed at $H_{c1} = 10.6$~T for the out-of-plane field configuration, $H\parallel c$. 
This feature was reproduced in several independent field pulses in both up- and down-ramping processes. 
These anomalies allow us to assign the field-induced Y, UUD, and V phases, together with low-field antiferromagnetic phases below 10~T, as schematically illustrated in Fig.~\ref{fig3}.

The plateau magnetization values of 0.66~$\mu_{\mathrm{B}}$ for $H\parallel ab$ and 0.76~$\mu_{\mathrm{B}}$ for $H\parallel c$, correspond to one-third of the saturation moments $gS\mu_\mathrm{B}$ for $S=1$, yielding effective estimates of $g_{ab}\simeq1.98$ and $g_c\simeq2.28$. 
We note that the absolute $g$ values inferred in this way depend on the magnetization normalization using low field PPMS or MPMS results, and here deviate from the Curie--Weiss results. 
Nevertheless, the larger magnetic response along the $c$ axis is consistently observed in measurements on different sample batches.
An independent determination of the $g$ factor, for example by ESR spectroscopy, would be valuable for establishing its absolute values.

The occurrence of the one-third magnetization plateau, corresponding to the UUD phase, has been well established in both classical~\cite{Kawamura1985} and quantum~\cite{Chubukov1991,Yamamoto2014} TLAF models. 
In particular, the evolution of the phase boundaries as a function of $1/S$ in the isotropic Heisenberg model has been investigated theoretically~\cite{Coletta2016}. 
Assuming an isotropic approximation and $S=1$ for $\mathrm{Rb_2Ni(SeO_3)_2}$, we adopt the theoretical values of $2.765J$ and $3.642J$ for the lower and upper critical fields of the UUD phase, respectively. 
This yields an estimate of the nearest-neighbor exchange coupling $J$ in the range of $0.8$--$1.4$~meV.
We note that the plateau width at zero temperature may differ from what we measured at 4.2~K, and a quantitative comparison should ideally be made at zero temperature.

\begin{figure}[bt]
\includegraphics[width=\columnwidth]{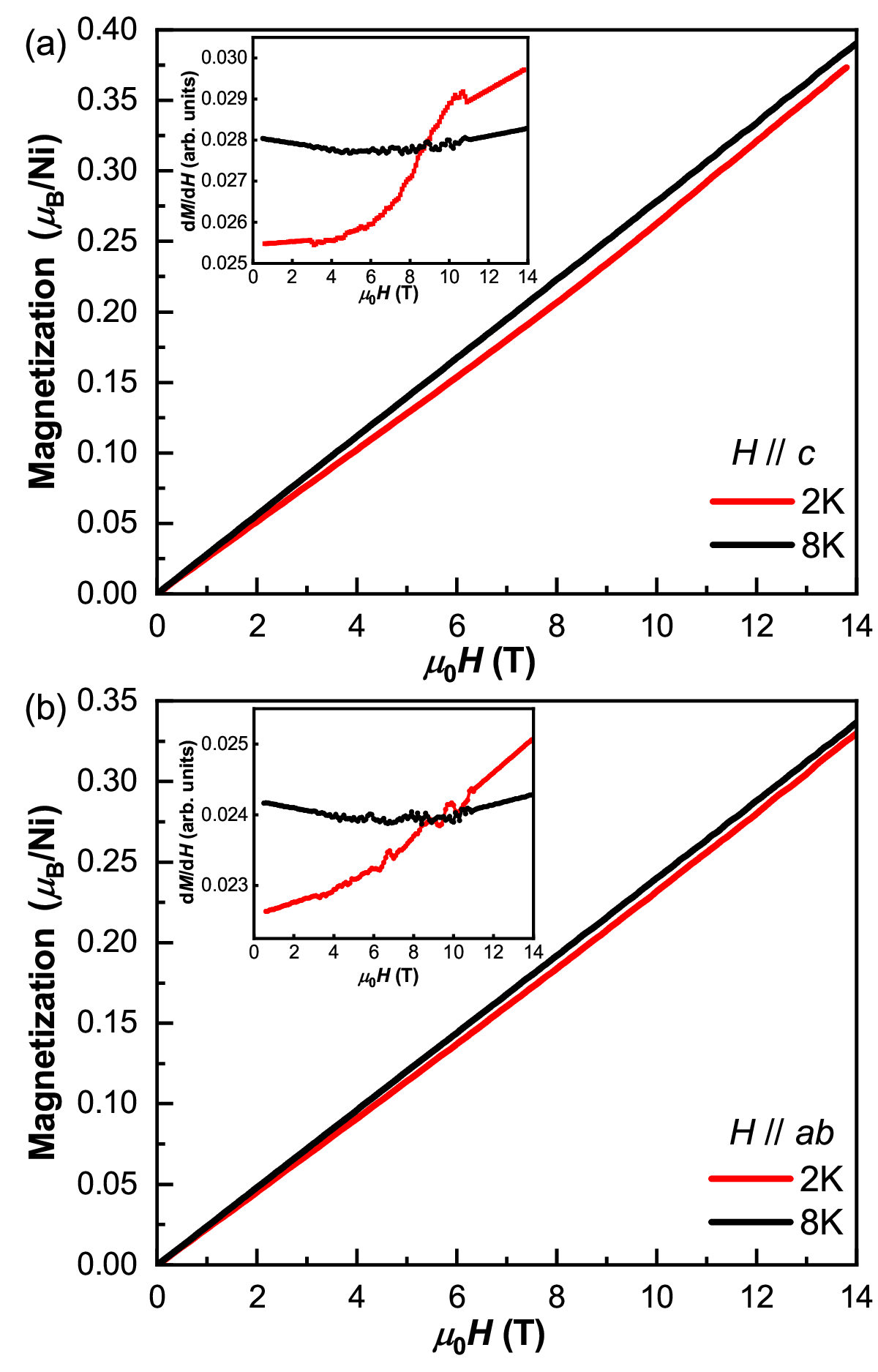}
\caption{(a) Out-of-plane ($H \parallel c$) and (b) in-plane ($H \parallel ab$) field dependent magnetization of Rb$_2$Ni(SeO$_3$)$_2$ at $T = 2$ (in red) and 8~K (in black). Insets show the $dM/dH$ curves with the same color coding.
}
\label{fig4}
\end{figure}

To examine the weak anomaly near $H_{c1}=10.6$~T in greater detail, we performed quasistatic magnetization measurements using the PPMS.
Representative data at 2~K in the AFM state and at 8~K in the paramagnetic state are shown in Fig.~\ref{fig4}. 
The magnetization is again larger for $H\parallel c$ than for $H\parallel ab$.
For $H\parallel c$ (Fig.~\ref{fig4} (a)), the 2~K magnetization curve exhibits a weak but discernible inflection, accompanied by an anomaly in $dM/dH$. 
For $H\parallel ab$ [Fig.~\ref{fig4}(b)], the corresponding feature is too weak to be clearly resolved.

\begin{figure*}[htb!]
\centering
\includegraphics[width=14cm, height=11.71cm]{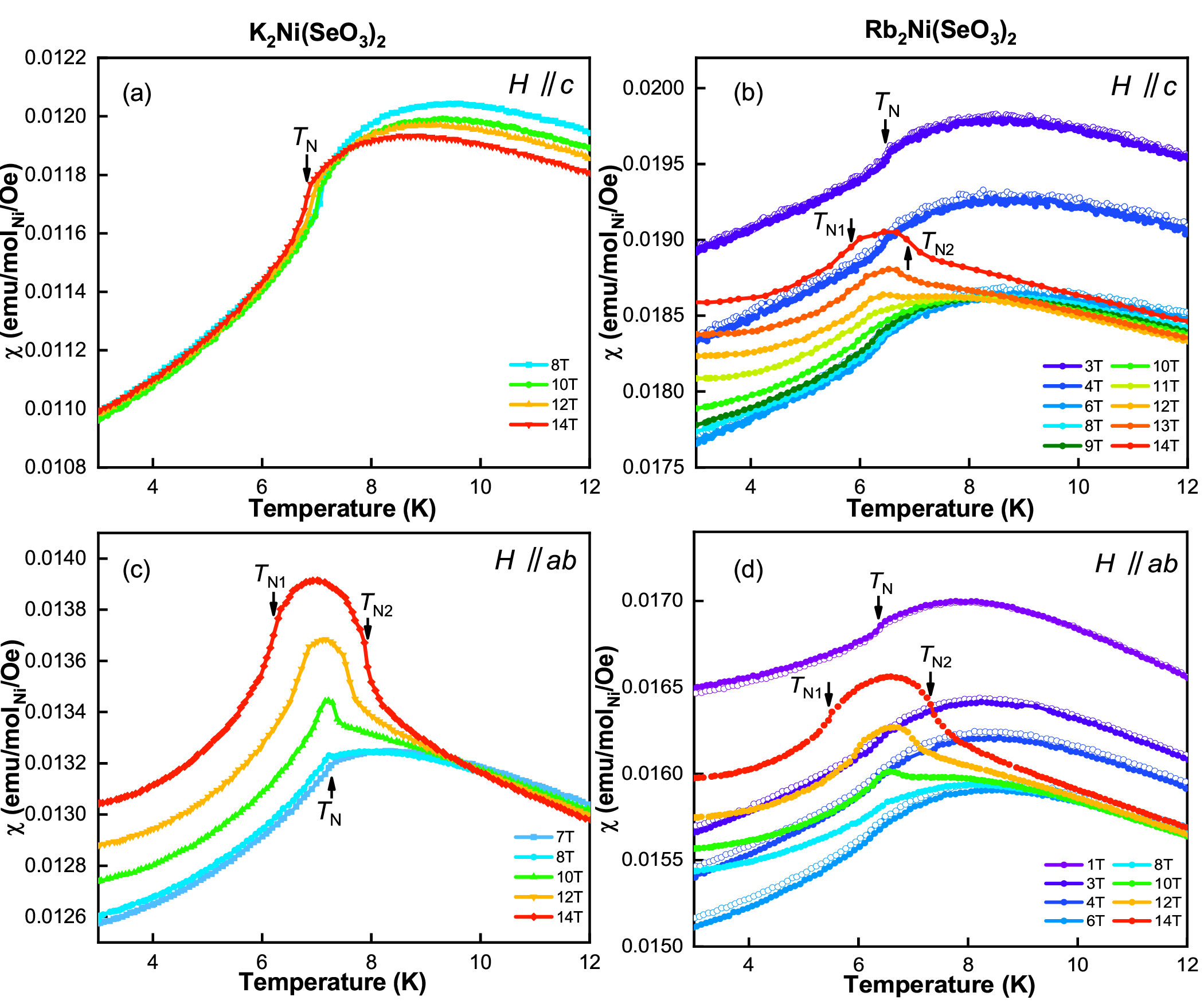}
\caption{Temperature dependence of the magnetic susceptibility for $\mathrm{K_2Ni(SeO_3)_2}$ (left panels) and $\mathrm{Rb_2Ni(SeO_3)_2}$ (right panels), with magnetic fields parallel to $c$ axis in panels (a-b) and parallel to the $ab$ plane in panels (c-d). The same color coding for the magnetic fields is used across all panels.
Field-cooled (FC) and zero-field-cooled (ZFC) data are denoted by solid and open symbols, respectively, with no significant differences between them. Arrows mark the transition temperatures determined from $\mathrm{d}\chi/\mathrm{d}T$. 
}
\label{fig5}
\end{figure*}

\begin{figure*}[htb!]
\centering
\includegraphics[width=14cm, height=12.25cm]{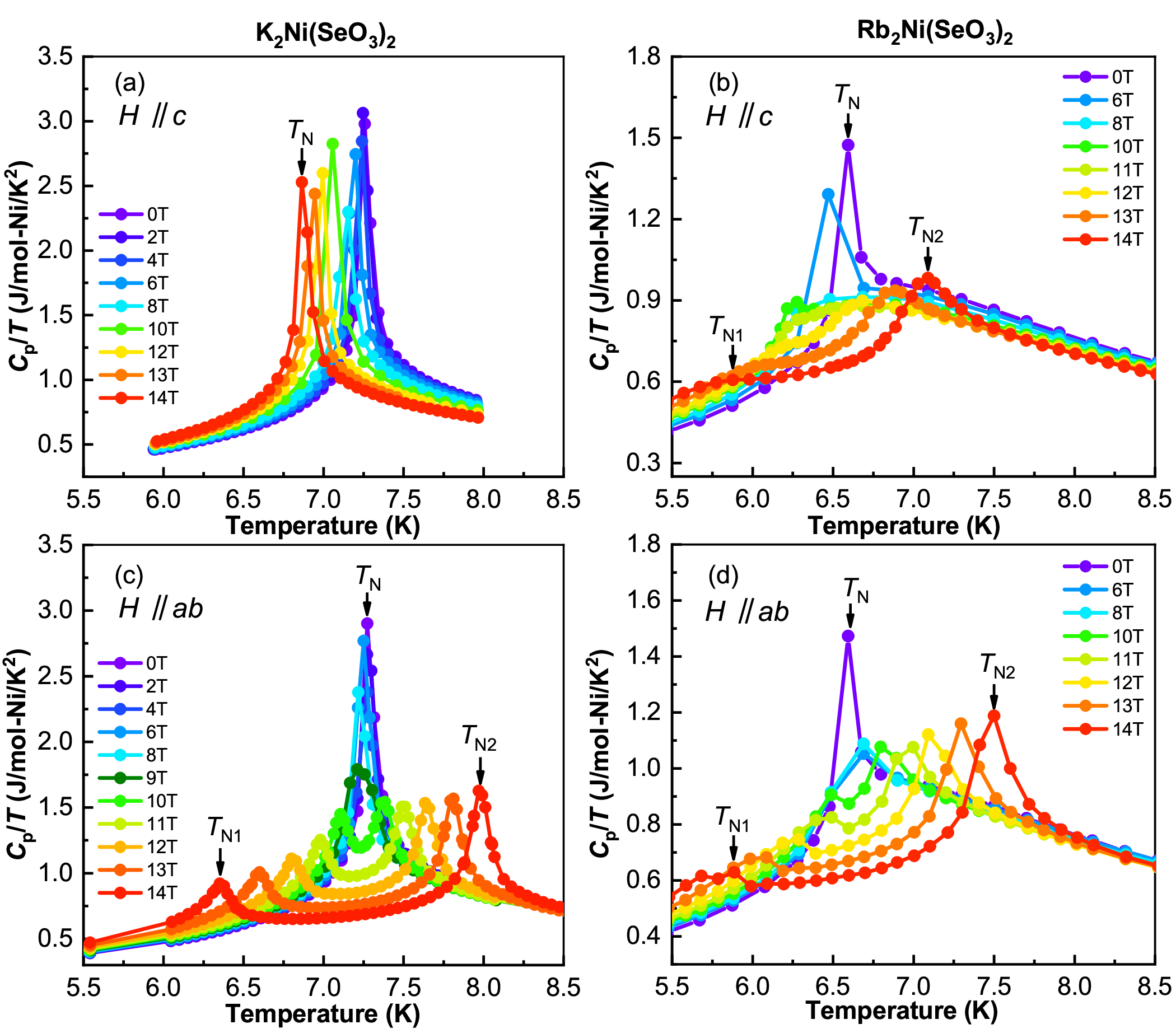}
\caption{Specific heat $C_p(T)$ of K$_2$Ni(SeO$_3$)$_2$ (left panels) and Rb$_2$Ni(SeO$_3$)$_2$ (right panels), with magnetic fields parallel to $c$ axis in panels (a-b) and parallel to the $ab$ plane in panels (c-d). The same color coding for the magnetic fields is used across all panels. Arrows mark the transition temperatures determined from the peaks in the curves.
}
\label{fig6}
\end{figure*}

The 10.6~T anomaly for $H\parallel c$ was not reported in $\mathrm{K_2Ni(SeO_3)_2}$, indicating different phase behavior between these sibling compounds. We further explore this by the temperature dependent magnetic susceptibility in Fig.~\ref{fig5} and the specific heat in Fig.~\ref{fig6}.

In Fig.~\ref{fig5}, we see AFM transitions as anomalies in the $\chi(T)$ curves. In $\mathrm{Rb_2Ni(SeO_3)_2}$, the low field anomaly, characteristic of the N\'eel temperature $T_\mathrm{N}$, evolves into a hump with two transition temperatures $T_\mathrm{N1}$ and $T_\mathrm{N2}$, roughly above 10~T. This split of $T_\mathrm{N}$ occurs for both $H\parallel c$ and $H\parallel ab$ in $\mathrm{Rb_2Ni(SeO_3)_2}$, but only for $H\parallel ab$ in $\mathrm{K_2Ni(SeO_3)_2}$. The reverse of magnetic anisotropy, i.e.~\ $\chi_c > \chi_{ab}$ in $\mathrm{Rb_2Ni(SeO_3)_2}$ and $\chi_{ab} > \chi_{c}$ in $\mathrm{K_2Ni(SeO_3)_2}$, is also evident in Fig.~\ref{fig5}.

The split of $T_\mathrm{N}$ above 10~T and the field-induced phase evolution are further confirmed in the specific heat measurement in Fig.~\ref{fig6}. Here, sharp peaks in the specific heat curves mark the AFM transitions. The contrast is again evident that the split of $T_\mathrm{N}$ is observed in Fig.~\ref{fig6} (b-d) but not in Fig.~\ref{fig6}(a). When the out-of-plane field $H\parallel c$ is applied in $\mathrm{K_2Ni(SeO_3)_2}$, $T_\mathrm{N}$ only shifts to low temperatures. It is also noticed that the specific heat anomaly is broader in $\mathrm{Rb_2Ni(SeO_3)_2}$ than in $\mathrm{K_2Ni(SeO_3)_2}$. This might be associated with the more quasi-two-dimensional character of $\mathrm{Rb_2Ni(SeO_3)_2}$, as suggested by its larger interlayer spacing.

\begin{figure}[htb!]
\includegraphics[width=\columnwidth]{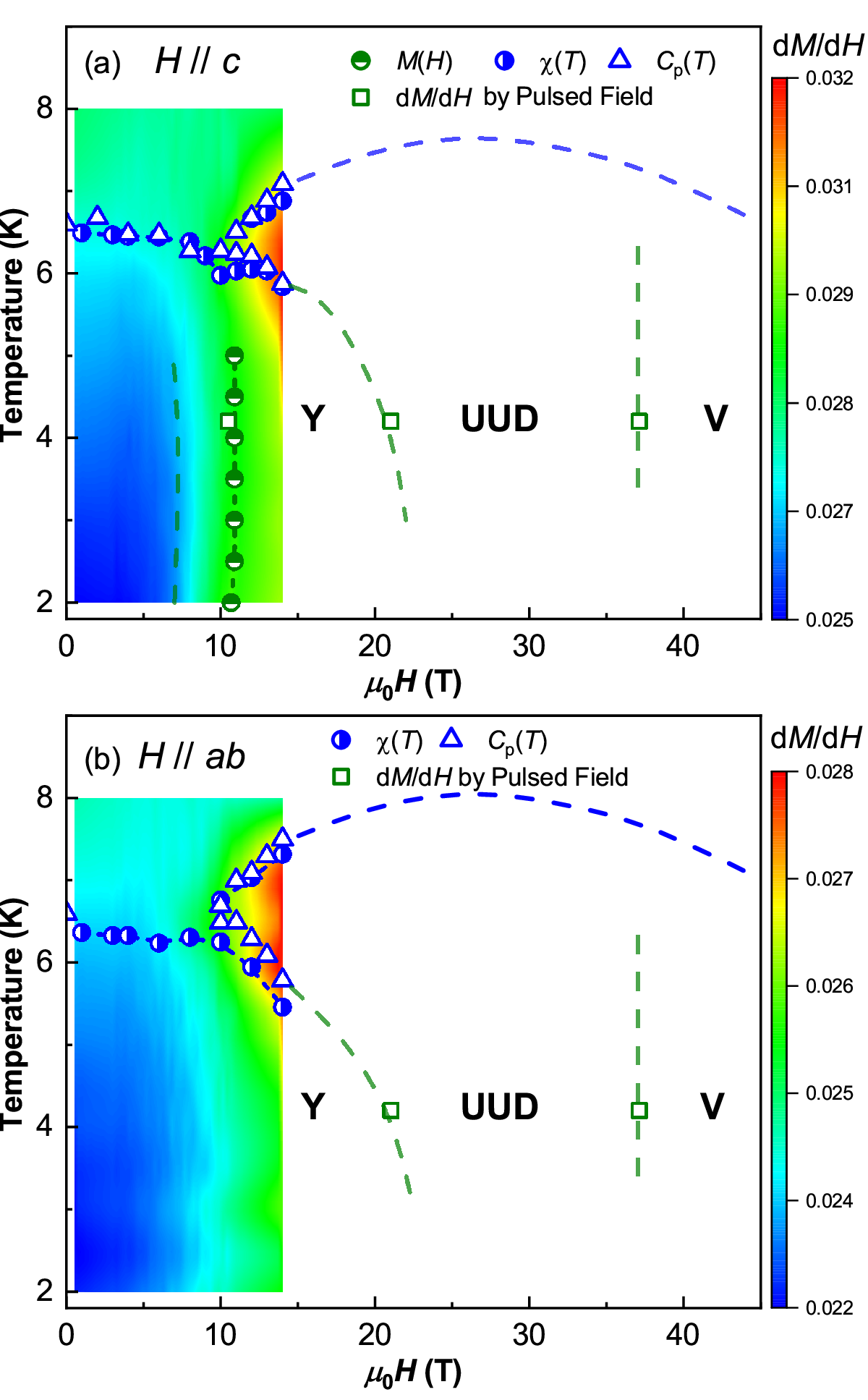}
\caption{Magnetic phase diagrams of $\mathrm{Rb_2Ni(SeO_3)_2}$ for (a) $H \parallel c$ and (b) $H \parallel ab$. Phase boundaries are determined from $M(H)$, $M(T)$, $C_{\mathrm{p}}(T)$ and pulsed high-field measurements. Above 14~T, the dashed lines serve as guides to the eye to indicate the evolution of the phase boundaries.
}
\label{fig7}
\end{figure}

Based on the combined results from magnetic susceptibility, specific-heat, and pulsed high-field measurements, we construct the magnetic phase diagrams of $\mathrm{Rb_2Ni(SeO_3)_2}$ for $H \parallel c$ and $H \parallel ab$, as shown in Fig.~\ref{fig7}. The color scale denotes the field derivative of the magnetization $dM/dH$ in PPMS measurements.
At fields below 10~T, we mark two low-field regions, for $H < 7~\mathrm{T}$ and $7~\mathrm{T} < H < 10~\mathrm{T}$, respectively.
An increase in $dM/dH$ is observed near 7~T. Above 10~T, we observe the split of the AFM transition temperature $T_{\mathrm{N}}$ and the anomaly at 10.6~T in the pulsed-field measurement. 
The two low-field phases are independently confirmed by nuclear magnetic resonance (NMR) spectra \cite{nmr-private}.
As the magnetic field increases further, the field-induced Y, UUD, and V phases emerge sequentially.
The phase diagrams are similar in both field orientations, indicating the system lies in the vicinity of the isotropic limit.

\section{Discussion}
The changes in magnetic anisotropy and the phase diagram upon replacing K with Rb are noteworthy, because the nonmagnetic ions are generally regarded as spacers between magnetic layers, with only a weak influence on magnetic properties.
Intriguingly, $\mathrm{Rb_2Ni(SeO_3)_2}$ resides in the easy-axis region, but has low-field anomalies preceding the field induced Y--UUD--V phases. This is not explained by the XXZ TLAF phase diagram with exchange anisotropy.
For example, for the easy-axis exchange anisotropy, Yamamoto et al.\ ~\cite{Yamamoto2014} predicted the field induced Y--UUD--V sequence, which has been realized experimentally in $\mathrm{K_2Co(SeO_3)_2}$ and $\mathrm{Na_2BaCo(PO_4)_2}$ (see Table~\ref{tab:materials}). But no low-field phases before such sequence has ever been reported. 
For the easy-plane exchange anisotropy, indeed a four-phase Umbrella--Y--UUD--V sequence is predicted in the slightly easy-plane regime in Yamamoto's phase diagram~\cite{Yamamoto2014} for $S = 1/2$, and a five-phase Umbrella--Fan--Y--UUD--V sequence in a more recent phase diagram for $S = 1$~\cite{Lv2026}. The $S=1/2$ compound $\mathrm{Ba_3CoSb_2O_9}$ and the $S=1$ compound $\mathrm{K_2Ni(SeO_3)_2}$ both exhibit easy-plane exchange anisotropy of a few percent (see Table~\ref{tab:materials}). However, such a multiple-transition sequence has not been observed in these easy-plane materials. Not to mention that the easy-axis anisotropy in  $\mathrm{Rb_2Ni(SeO_3)_2}$ seems to conflict with the Umbrella phase, which is evolved from the planar 120$^\circ$ configuration in the easy-plane regime.
Thus, the field induced phases in $\mathrm{Rb_2Ni(SeO_3)_2}$ cannot be fully accounted for by theoretical predictions when only exchange anisotropy is considered.

We note that $\mathrm{Rb_2Ni(SeO_3)_2}$ lies close to the isotropic limit. Weak interactions such as interlayer coupling could be important in selecting the ground state out of the classical degenerate manifolds~\cite{Woodland2025}.
We don't exclude such possibility, but consider the effects of turning on different sources of anisotropies, i.e.\ the exchange anisotropy and single-ion anisotropy. The Hamiltonian can be written as
\begin{align*}
\hat{H} = J
&\sum_{\langle i,j\rangle}
(S_i^x S_j^x + S_i^y S_j^y + \Delta S_i^z S_j^z)
\\
&- D\, \sum_i (S_i^z)^2
- \mu_{\mathrm B}\sum_i
\mathbf{B}\cdot\mathbf{g}\cdot\mathbf{S}_i,
\end{align*}
where $\Delta$ characterizes the exchange anisotropy, and $D$ denotes the single-ion anisotropy. 
Both single-ion and exchange anisotropies can arise from spin--orbit coupling (SOC).
Single-ion anisotropy originates from SOC on the local crystal-field environment, which favors local spin orientations.
Exchange anisotropy arises from SOC on exchange pathways and favors particular orientations of collective spin alignment.
These tendencies can compete.
For example, easy-plane exchange anisotropy may favor transverse spin alignment and enhance quantum fluctuations in a system with easy-axis single-ion anisotropy.
Although changes in exchange anisotropy $\Delta$ and single-ion anisotropy $D$ between $\mathrm{K_2Ni(SeO_3)_2}$ and $\mathrm{Rb_2Ni(SeO_3)_2}$ cannot yet be disentangled, we explore their possible combined effects.

\begin{table*}[t]
\caption{\label{tab:materials}
Representative triangular-lattice antiferromagnets. The exchange coupling $J$ and single-ion anisotropy $D$ are given in meV, and are defined in the text. Here $\Delta >1$, and $D>0$ denotes easy-axis anisotropy.
}
\centering

\begin{tabular}{ @{} l l c c c c c c @{} }
\toprule

Anisotropy & Compound & $S$ & $J$ & $\Delta$ & $D$ & Phase evolution & Refs.
\\
\midrule

\multirow{3}{*}{Easy-Axis}
& $\mathrm{K_2Co(SeO_3)_2}$ & $1/2$ & $0.21$ & $14$ & --
& $H\parallel c$: Y--UUD--V
& \cite{Zhu2024,Chen2026}
\\
\cmidrule(lr){2-8}

& $\mathrm{Na_2BaCo(PO_4)_2}$ & $1/2$ & $0.076$ & $1.645$ & --
& $H\parallel c$: Y--UUD--V
& \cite{Sheng2022,Sheng2025}
\\
\cmidrule(lr){2-8}

& $\mathrm{Na_2BaNi(PO_4)_2}$ & $1$ & $0.032$ & $1.1$ & $0.127$
& Spin-nematic phases
& \cite{sheng2025bose,ShengPRL2025,Huang2025}
\\

\midrule

\multirow{2}{*}{Easy-Plane}
& \multirow{2}{*}{$\mathrm{Ba_3CoSb_2O_9}$}  
& \multirow{2}{*}{$1/2$} 
& \multirow{2}{*}{$1.7$} 
& \multirow{2}{*}{$0.85$--$0.95$} 
& \multirow{2}{*}{--}
& $H\parallel c$: Umbrella--V
& \multirow{2}{*}{\cite{Ito2017,Kamiya2018,Macdougal2020}} 
\\

& & & &  &
& $H\parallel ab$: Y--UUD--V
&  
\\
\midrule

\multirow{2}{*}{Modified}
& $\mathrm{K_2Ni(SeO_3)_2}$ & $1$ & $1.1$ & $0.94$ & --
& $H\parallel c$: no transition up to 14~T
& \cite{Kim2026}
\\

\cmidrule(lr){2-8}

&$\mathrm{Rb_2Ni(SeO_3)_2}$ & $1$ & -- & -- & --
& low-field-phases--Y--UUD--V
& This work
\\

\bottomrule
\end{tabular}
\end{table*}

When both $\Delta$ and $D$ are considered, additional phases may emerge.
A recent study~\cite{Seifert2022} demonstrated that while easy-axis exchange favors the coplanar Y phase, adding an easy-plane single-ion term can favor a modified Y phase by lying one sublattice spin in the plane (see Fig.~6 and the `zAFM3' phase in Ref.~\cite{Seifert2022}). If we denote it as a sideways Y or `sY', our on-going simulations suggest that a sY--Y--UUD--V transition sequence is possible for out-of-plane magnetic field in this region. 
On the other hand, when the exchange interaction is easy-plane like and the single-ion anisotropy is easy-axis like, we might anticipate Umbrella phase at low field from exchange anisotropy, while single-ion anisotropy is responsible for a larger $g_c$ value.
Then we could have the field induced Umbrella--Y--UUD--V sequence, even when the overall magnetic anisotropy is easy-axis.
The proximity to the Heisenberg limit in $\mathrm{Rb_2Ni(SeO_3)_2}$ implies compensation between exchange and single-ion anisotropies. 
The interplay of these anisotropies could generate rich physics that remains to be explored.

Very recently, the quadrupolar phases in the $S = 1$ system have attracted a lot of attention~\cite{sheng2025bose}. Such phases become possible when both bilinear exchange coupling and the single-ion term are present. Thus, including the single-ion term could bring not only new dipolar orders but also quadrupolar orders. One example is a `Q-up-down' phase predicted for easy-axis exchange and easy-plane single-ion anisotropies for $S=1$ TLAF ~\cite{Seifert2022}. 
Recently, the two-magnon bound state and spin-nematic phases have been demonstrated in the $S = 1$ compound $\mathrm{Na_2BaNi(PO_4)_2}$ for large easy-axis-type $D/J$~\cite{sheng2025bose} (see Table.~\ref{tab:materials}). 
It is tempting to observe quadrupolar physics in systems with much larger $J$, as it would facilitate the experimental resolution of the exotic spin excitations. $\mathrm{Rb_2Ni(SeO_3)_2}$ has approximately 50-fold larger exchange parameter than $\mathrm{Na_2BaNi(PO_4)_2}$ does. Although it remains unclear whether similar physics, which requires at least 2--4~meV of $D$, can be realized here, other forms of quadrapolar phases are possible and we highlight that the single-ion anisotropy in the $S=1$ system opens such opportunities. These remain to be investigated, and precise determination of the interaction parameters in $\mathrm{Rb_2Ni(SeO_3)_2}$ is desired.

\section{Conclusion}
In conclusion, we have investigated the magnetic properties of the $S=1$
triangular-lattice antiferromagnet $\mathrm{Rb_2Ni(SeO_3)_2}$. Compared with the isostructural compound $\mathrm{K_2Ni(SeO_3)_2}$, $\mathrm{Rb_2Ni(SeO_3)_2}$ exhibits a reversed susceptibility anisotropy and distinct field-induced phase behavior. A one-third magnetization plateau is observed for both in-plane and out-of-plane magnetic fields, with comparable plateau boundaries. Additional low-field anomalies and a split of the N\'eel transition near 10~T further distinguish its magnetic phase diagram from that of $\mathrm{K_2Ni(SeO_3)_2}$.
Our observation demonstrates that even subtle tuning can produce measurable effects in a nearly isotropic triangular-lattice magnet, and suggests anisotropy competition as a promising path towards novel phases. 
The interplay between exchange anisotropy and single-ion anisotropy could induce exotic quantum phases in TLAF systems, which remains to be explored both theoretically and experimentally. 

\section*{Acknowledgement}
This work was supported by the National Key R{\&}D Program of China (Grant No.~2023YFA1406500), the National Natural Science Foundation of China (Grant Nos.~12304185 and 12574157), the Guangdong Basic and Applied Basic Research Foundation (Grant No.~2026B1515020020), and the Guangdong Provincial Quantum Science Strategic Initiative (Grant No.~GDZX2401007).
Xuejuan Gui acknowledges support from the Joint PhD Program scholarship funded by Renmin University of China.


\bibliographystyle{apsrev4-2}
\bibliography{RbNi212}

\end{document}